\documentclass[sigconf,10pt,prologue,dvipsames,numbers,sort&compress]{acmart}

\renewcommand\footnotetextcopyrightpermission[1]{} 
\usepackage{comment}
\usepackage{enumerate}
\usepackage{color}
\usepackage{xcolor}
\usepackage{relsize}
\usepackage{graphicx}
\usepackage{subfigure}
\usepackage{tabularx}
\usepackage{booktabs}
\usepackage{listings}
\usepackage{amssymb}
\usepackage{wrapfig}
\usepackage{algorithm}
\usepackage{algpseudocode}
\usepackage[utf8]{inputenc}
\usepackage[english]{babel}
\usepackage{amsthm}

\usepackage{mathtools}
\usepackage{lipsum} 
\usepackage{CJKutf8}
\usepackage{listings}
\usepackage{algpseudocode}
\usepackage{algorithmicx,algorithm}

\hypersetup{
	colorlinks   = true,
	citecolor    = cyan,
	linkcolor    = magenta,
	urlcolor	= blue
}
\definecolor{cobalt}{rgb}{0.0, 0.28, 0.67}

\newcommand{\xin}[1]{\textcolor{cobalt}{Xinying: #1}}

\newcommand{\figspace}{\vspace{-1mm}}

\definecolor{codegreen}{rgb}{0,0.6,0}
\definecolor{codegray}{rgb}{0.5,0.5,0.5}
\definecolor{codepurple}{rgb}{0.58,0,0.82}
\definecolor{backcolour}{rgb}{0.95,0.95,0.92}

\lstdefinestyle{mystyle}{
    frame = single,
  commentstyle=\color{codegreen},
  keywordstyle=\color{magenta},
  numberstyle=\tiny\color{codegray},
  stringstyle=\color{codepurple},
  basicstyle=\footnotesize,
  breakatwhitespace=false,         
  breaklines=true,                 
  captionpos=b,                    
  keepspaces=true,                 
  numbers=left,                    
  numbersep=5pt,                  
  showspaces=false,                
  showstringspaces=false,
  showtabs=false,                  
  tabsize=2
}

\begin{document}
\begin{CJK}{UTF8}{gbsn}

\title{Distributed Nonblocking Commit Protocols for Many-Party Cross-Blockchain Transactions}

\author{Xinying Wang, Olamide Timothy Tawose, Feng Yan, Dongfang Zhao}
\affiliation{%
  \institution{University of Nevada, Reno}
}

\begin{abstract}
The interoperability across multiple blockchains would play a critical role in future blockchain-based data management paradigm.
Existing techniques either work only for two blockchains or requires a centralized component to govern the cross-blockchain transaction execution,
neither of which would meet the scalability requirement.
This paper proposes a new distributed commit protocol, namely \textit{cross-blockchain transaction} (CBT), 
for conducting transactions across an arbitrary number of blockchains without any centralized component.
The key idea of CBT is to extend the two-phase commit protocol with a heartbeat mechanism to ensure the liveness of CBT without introducing additional nodes or blockchains.
We have implemented CBT and compared it to the state-of-the-art protocols, demonstrating CBT's low overhead (3.6\% between two blockchains, less than $1\%$ among 32 or more blockchains) and high scalability (linear scalability on up to 64-blockchain transactions).
In addition, we developed a graphic user interface for users to virtually monitor the status of the cross-blockchain transactions.
\end{abstract}

\maketitle

\figspace
\section{Introduction}

A blockchain offers an immutable, decentralized, and anonymous mechanism for transactions between two users on the same blockchain. 
Blockchain was not originally designed for online transactional processing (OLTP) workloads;
instead, it aimed to offer an autonomous and temper-proof ledger service among mutually-distrusted parties and therefore, 
early blockchain systems can deliver only mediocre transaction throughput.

One natural question to ask is whether and how can we adopt blockchains to efficiently handle OLTP workloads such that both autonomy and performance can be simultaneously achieved. 
Indeed, most of recent works focus on this direction.
For instance, in~\cite{mhindi_vldb19,jwang_nsdi19}, authors advocate to leverage blockchains for OLTP workloads with various optimizations (e.g., sharding~\cite{hdang_sigmod19}, sidechains~\cite{sidechain}) to boost up the transaction throughput of blockchains,
such that blockchains would deliver similarly high performance as relational databases and become a competitive alternative to the latter as a general-purpose data management system.

There is yet another critical issue that must be addressed before blockchains can be widely adopted as a general data management system:
the interoperability across heterogeneous blockchains.
While SQL is available between distinct relational database implementations, 
no such standardization or interface exists for blockchains.
Even worse, there is no known \textit{mechanism} about how to transfer data among multiple blockchains without a central exchange.
Recent attempts (e.g., Cosmos~\cite{cosmos}) on such cross-blockchain transactions are all \textit{ad hoc}:
making strong assumptions on the blockchains such as their consensus protocols and programming interface.
In addition, existing cross-blockchain approaches exhibit various limitations such as limited scalability and significant performance overhead.
To make it more specific, what follows lists four outstanding limitations exhibited by state-of-the-art cross-blockchain solutions:

\textbf{(1) Centralized Broker.} The transactions between heterogeneous blockchains are managed by a third-party, usually implemented as another blockchain (it is called a \textit{hub} in Cosmos).
This is against the decentralization principle of blockchains:
the broker would become a performance bottleneck, an attack target, and a single-point-of-failure.
Similarly, a recent work called AC3~\cite{vzakhary_arxiv19} employs an extra component (known as \textit{witness blockchain}) as a central authority to govern the cross-chain operations.
    
\textbf{(2) Two-Party Transactions.} The protocols used by existing cross-blockchain systems are derived from the \textit{sidechain protocol}~\cite{sidechain},
    which was originally designed for transferring assets between Bitcoin~\cite{bitcoin} and another cryptocurrency.
    The sidechain protocol speaks of nothing about three- or multi-party transactions;
    in fact, Cosmos only supports transferring assets between Bitcoin~\cite{bitcoin} and Ethereum~\cite{ethereum}.
A more recent line of works~\cite{maurice_vldb19,mherlihy_podc18} are based on two-party Atomic Cross-Chain Swaps (ACS);
however, ACS cannot guarantee the atomicity of the multi-chain deals as a whole.

\textbf{(3) Performance.} 
The sidechain protocol~\cite{sidechain} took hours, if not days, to commit a cross-blockchain transaction. 
    The main reason for this is due to the possible branches from the participating blockchains.
    In any participating blockchain,
    only one (i.e., the longest one) branch will remain valid and any transactions from the shorter branches will rollback.
    This is not a problem if all of the transaction parties are from the same blockchain;
    But for cross-blockchain transactions,
    they must wait for the (longest) branch to stand out.
    
\textbf{(4) Interface.} 
The centralized broker requires the users to pack their cross-blockchain transactions with the provided interface.
    It would create portability issues when the users concurrently work with multiple cross-blockchain platforms.
    What we need is a common interface with which different blockchains (and their users) can communicate.
    SQL is an excellent example for relational databases.

\begin{figure}[!t]
	\centering
	\includegraphics[width=70mm]{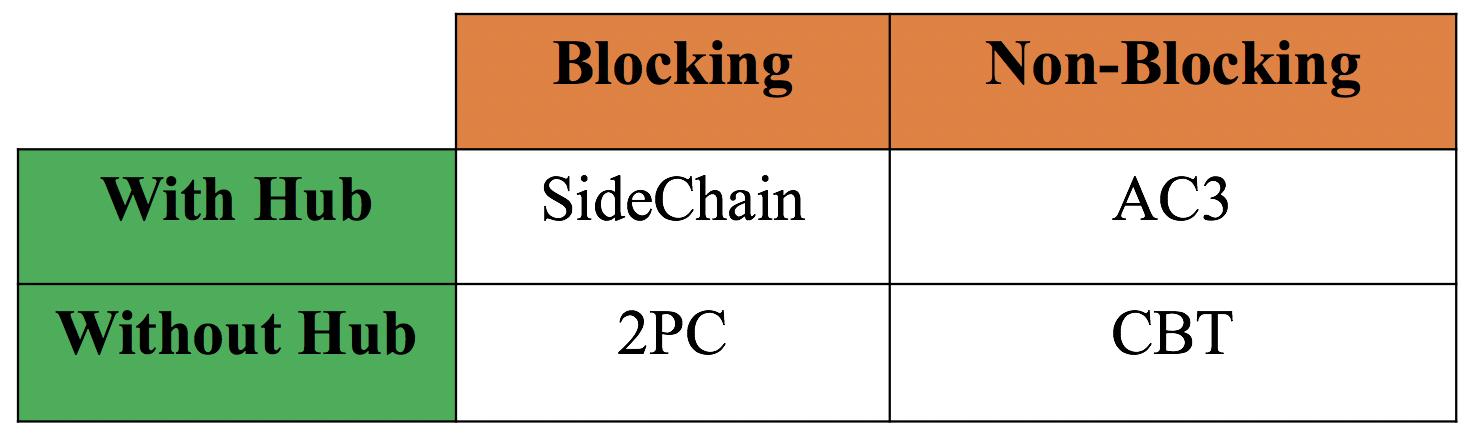}
	\caption{Feature comparison among cross-blockchain transaction protocols.}
	\label{fig:chart}
	
\end{figure}

Figure~\ref{fig:chart} summarizes candidate solutions along with two of the most important features in the realm of cross-blockchain transactions.
As we can see, existing works are limited to centralized design (i.e., the requirement of a hub), or the potential blocking of the operation, or both.
To this end, we propose a new protocol, namely cross-blockchain transaction (CBT), 
to overcome the above limitations.
The key design objectives of CBT is to eliminate the centralized component and to ensure liveness (i.e., nonblocking).

The idea of decentralized and nonblocking cross-blockchain transactions was recently presented at~\cite{dzhao_cidr20}, as an abstract.
This demo paper shares recent progress on the following perspectives:
\begin{itemize}
    \item We designed a set of protocols, i.e., CBT, that enables a transaction completed (or, aborted altogether) across an arbitrary number of blockchains;
    \item We implemented CBT on BlockLite~\cite{xwang_cloud19} and validated the effectiveness of CBT over popular blockchain implementations;
    \item We evaluated CBT against multiple state-of-the-art protocols for multi-party transactions in blockchains;
    \item We developed a graphic interface allowing non-expert users to get hands-on experience on and concrete understanding of cross-blockchain transactions.
\end{itemize}

In the remainder of this paper, we will formulate the protocol design of CBT in~\S\ref{sec:protocol},
detail CBT's implementation and report the experimental results in~\S\ref{sec:result},
and finally demonstrate the graphic user interface of CBT and other candidate protocols in~\S\ref{sec:demo}.

\section{Cross-Blockchain Transactions}
\label{sec:protocol}

\subsection{An Example with Three Blockchains}

\begin{figure*}[!t]
    \centering
    \includegraphics[width=170mm]{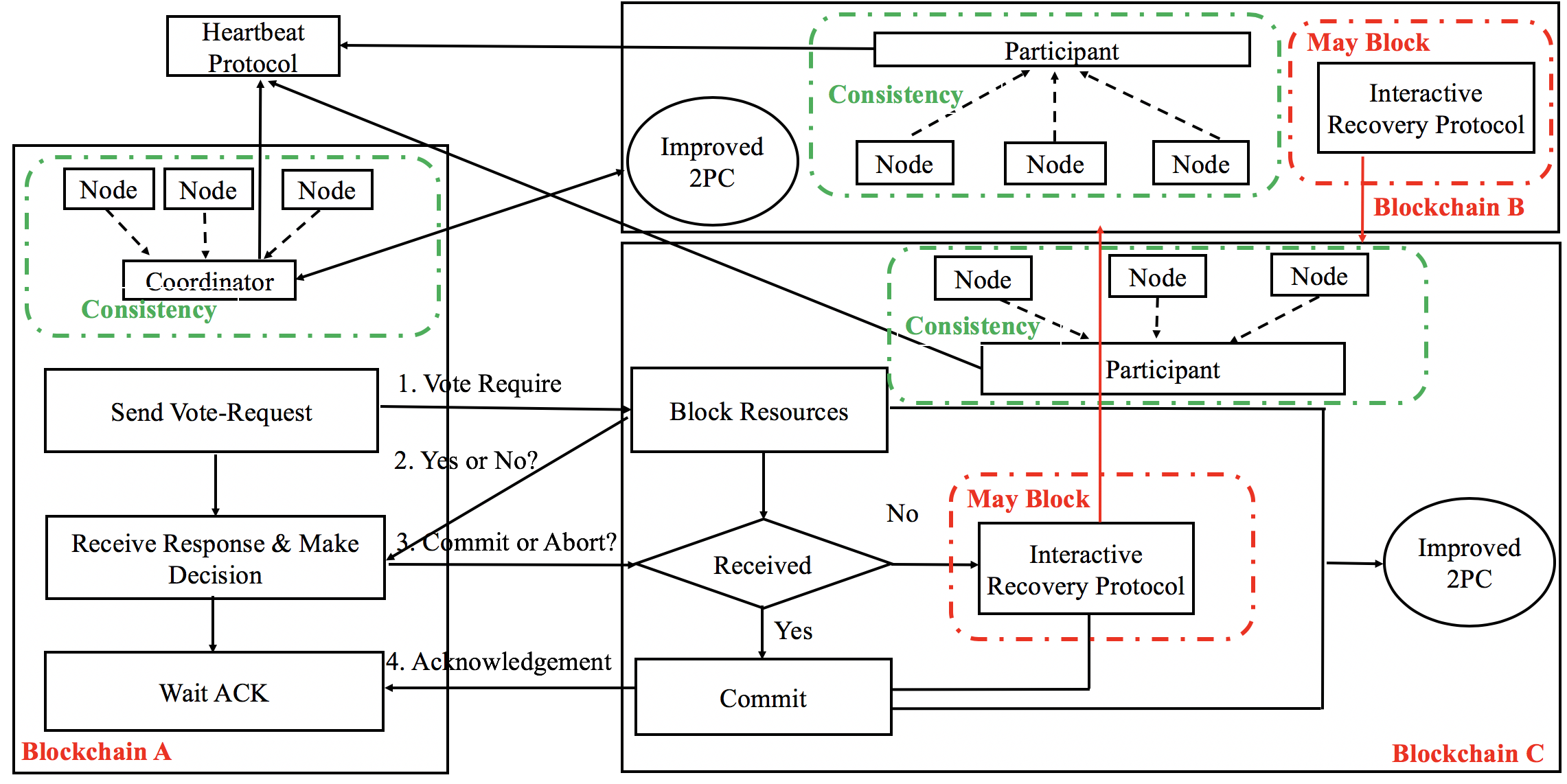}
    \caption{A simplified running example of CBT across three blockchains: \textit{A}, \textit{B}, and \textit{C}.}
    \label{fig:2pc}
    \figspace    
    \figspace
\end{figure*}

Before presenting the formal protocols, we illustrate the execution of CBT on a 3-blockchain scenario.
It should be noted that the protocol can be applied to an arbitrary number of blockchains;
we chose three because the 2-blockchain case is trivial and 3-blockchain is reasonably understandable for users who are interested in generalizing the protocol into a $n$-blockchain case, where $n \in \mathbb{Z}^+$ and $n > 1$.

The overall flow is illustrated in Figure~\ref{fig:2pc} where three blockchains $A$, $B$, and $C$ cooperatively complete a transaction.
We make the following assumptions and define some notations.

\begin{enumerate}

\item We assume one of the blockchains is elected as the \textit{coordinator} based on some election algorithms. Let's say $A$ is the coordinator, $B$ and $C$ are the participants. 
If both $B$ and $C$ agree on the transaction, 
the transaction will be committed; Otherwise, the proposed transaction will be rollback.

\item An \textit{uncertain period} is the period from the point when the participant votes \textit{Yes} to the point it gets enough information to make a final decision.

\item \textit{Blocking State}: during the uncertain period, if the coordinator's primary node crashes, and it only sends a final decision (COMMIT or ABORT) to some participants, 
then the other participants will be blocked until a new primary node is selected and online.

\item \textit{DT log}: each blockchain maintains a distributed transaction log (DT log), 
in which the coordinator and participants keep a write-ahead log of the transaction. 
The DT log must be stored in a persistent storage.

\item \textit{Heartbeat}: a heartbeat protocol is employed by each blockchain. 
Specifically, if a worker node does not receive a reply from the primary node within the heartbeat time, the primary node will be replaced (by another live worker node) and restarted as a worker.
\end{enumerate}

The normal execution (i.e., no failures) works as follows,
pretty much similar to 2PC but among the representative nodes from distinct blockchains:
\begin{enumerate}
\item[1.] The Commit Request Phase
    \begin{enumerate}
    \item[(1.1)] The coordinator sends a VOTE-REQUEST to all participants.
    \item[(1.2)] When the participant receives the VOTE-REQUEST, it will respond YES or NO to the coordinator.
    If the participant votes NO, it can unilaterally terminate the protocol. 
    \end{enumerate}
\item[2.]The Commit Phase
    \begin{enumerate}
    \item[(2.1)] The coordinator collects votes from all participants. 
    If all the responses are YES, then the coordinator sends a COMMIT message to all participants; Otherwise, the coordinator aborts and sends an ABORT message to all participants who have voted YES.
    \item[(2.2)] Each participant who votes YES needs to wait for a COMMIT or ABORT message from the coordinator. After receiving the message, the participant performs the corresponding action and terminates.
    \end{enumerate}
\end{enumerate}

The above 2PC protocol will only work in normal cases and will block in some scenarios. 
For instance, there will be cases where the message cannot be delivered due to the network errors or a participant is offline from a specific blockchain. 
To this end, we propose Improved 2PC (detailed in Algorithms 1 and 2) comprising a heartbeat monitoring mechanism (detailed in Algorithm 7). 
The exception and its handling is as follows (the step numbers refer to the arrows between blockchains $A$ and $C$ in Figure~\ref{fig:2pc}): 

\begin{enumerate}

\item In step 2, the participant waits for a VOTE-REQUEST from the coordinator. 
Let's assume that any participant can unilaterally decide to abort before it votes YES. 
Therefore, if there is a timeout when the participant is waiting for a VOTE-REQUEST, 
it can unilaterally choose to abort and stop the communication.

\item In step 3, the coordinator needs to wait for responses (YES or NO) from all participants. 
The coordinator can unilaterally decide whether to abort, but an ABORT message must be sent to each participant who previously sends a YES message.

\item In step 4, the participant who has responded YES is in an uncertain period when it is waiting for the final decision (COMMIT or ABORT) from the coordinator. 
Unlike previous two cases where blockchain can make decisions unilaterally, if a timeout occurs in this case, the participant needs to consult with other participants to decide the next action, which warrants a call to the Interactve Recvery Protocol (detailed in Algorithms 3 and 4).
\end{enumerate}

The Interactive Recovery Protocol works as follows.
If participant \textit{p} times out in the uncertainty period, 
it will send a DECISION-REQUIRE message to all the other participants, whose set is indicated by $q$, and ask if any member in $q$ knows the outcome of the DECISION or if it can make a DECISION unilaterally. 
Assuming $p$ is the initiator and $q$ is the responder. There are four cases (2 and 3 can be grouped together) as follows:

\begin{enumerate}

\item $q$ decides to commit (either as a coordinator or a participant): 
$q$ sends a COMMIT message to $p$, and then $p$ does the same;

\item $q$ decides to abort (either as a coordinator or a participant): 
$q$ sends an ABORT message to $p$, and $p$ does the same;

\item $q$ is a participant. $q$ unilaterally decides to abort even though $q$ has not voted. 
It then sends an ABORT message to $p$, and $p$ does the same;

\item $q$ is a participant. $q$ has voted YES and is also in an uncertain period.
It cannot help $p$ to reach a decision, and $p$ does nothing.
\end{enumerate}

In addition, the internal consistency of a blockchain itself needs to be maintained since each blockchain is composed of multiple nodes. 
The protocols are detailed in Algorithms 5 and 6.

\subsection{Protocols}

Most protocols are self-explanatory.
We skip the proof of correctness and complexity analyses due to limited space.


\begin{algorithm}
    \caption{Improved 2PC Algorithm (Coordinator)}
    \begin{algorithmic}[1] 
        \State send VOTE-REQUEST to all participants
        \State wait for vote messages from all participants
        \State on timeout:
            \State \qquad $P$ the set of processes from which YES was received
            \State \qquad write abort record in DT log
            \State \qquad send ABORT to all processes in $P$
            \If{all votes are YES and Coordinator votes YES}
                \State write commit record in DT log
                \State send COMMIT to all participants
            \Else 
                \State write abort record in DT log
                \State send ABORT to all processes in $P$
            \EndIf
    \end{algorithmic}
\end{algorithm}

\figspace

\begin{algorithm}
    \caption{Improved 2PC Algorithm (Participant)}
    \begin{algorithmic}[1] 
        \State wait for VOTE-REQUEST from coordinator
        \State on timeout
            \State \qquad write abort record in DT log
        \If{participant votes yes}
            \State write a yes record in DT log
            \State send YES to coordinator
            \State wait for decision message from coordinator
            \If{ VOTE-REQUEST from new coordinator}
                \If{decision message is COMMIT}
                    \State  write COMMIT record in DT log
                \Else
                    \State  write ABORT record in DT log
                \EndIf
            \Else
                \State on timeout initiate ask others protocol
            \EndIf
        \EndIf
    \end{algorithmic}
\end{algorithm}

\figspace

\begin{algorithm}
    \caption{Interactive Recovery Protocol (Initiator)}
    \begin{algorithmic}[1] 
        \State send DECISION-REQUIRE to all blockchains
        \State wait for decision message from any blockchains
        \State on timeout
            \State write ABORT record in DT log
        \If{decision message is COMMIT }
            \State write COMMIT record in DT log
        \Else
            \State write ABORT record in DT log
        \EndIf
    \end{algorithmic}
\end{algorithm}

\figspace

\begin{algorithm}
    \caption{Interactive Recovery Algorithm (Responder)}
    \begin{algorithmic}[1] 
        \State wait for DECISION-REQUIRE from any blockchains \textit{p}
        \If{responder has not voted, but decides to abort}
            \State send ABORT to \textit{p}
        \ElsIf{responder has decided to commit}
            \State send COMMIT to \textit{p}
        \ElsIf{responder has decided to abort}
            \State send ABORT to \textit{p}
        \Else 
            \State skip
        \EndIf
    \end{algorithmic}
\end{algorithm}
\figspace
\begin{algorithm}
    \caption{Consistency Algorithm (Leader)}
    \begin{algorithmic}[1] 
        \State write message to a block
        \State send message to all followers in cluster
        \If{crashes}
            \State restart as a follower
        \EndIf
    \end{algorithmic}
\end{algorithm}
\figspace
\begin{algorithm}
    \caption{Consistency Algorithm (Follower)}
    \begin{algorithmic}[1] 
        \State receive message from the leader
        \State write message to log
        \If{leader crashes}
            \State run for the new leader
            \If{elected}
                \State synchronize the log
            \Else 
                \State skip
            \EndIf
        \EndIf
    \end{algorithmic}
\end{algorithm}

\figspace

\begin{algorithm}
    \caption{Heartbeat Algorithm}
    \begin{algorithmic}[1] 
        \State build heartbeat thread
        \State encapsulate heartbeat requests
        \If{responses are normal}
            \State increment success counter by 1
        \Else 
            \State increment failure Counter by 1
        \EndIf    
        \If{success counter equals 3}
            \State reset success counter
        \ElsIf{failure counter equals 3}
            \State initiate a vote for a new leader
        \EndIf
    \end{algorithmic}
\end{algorithm}

\section{Implementation and Evaluation}
\label{sec:result}

We have implemented the proposed CBT protocol as well as two baseline protocols, i.e., 2PC~\cite{2pc} and AC3~\cite{vzakhary_arxiv19}, on the BlockLite system~\cite{xwang_cloud19}.
The source code will be accessible at:~\url{https://github.com/hpdic/cbt}.
The source code is written with Java of JDK 1.7.
The code base comprises about 5,190 lines of code. 
The code base has three major components: 
(i) the blockchain implementation including protocols and utilities; 
(ii) the network component including the communications among coordinator and participants;
and (iii) the graphic user interface developed with Java Swing.

\begin{figure}[!t]
	\centering
\subfigure[2PC Coordinator]{
	\includegraphics[width=40mm]{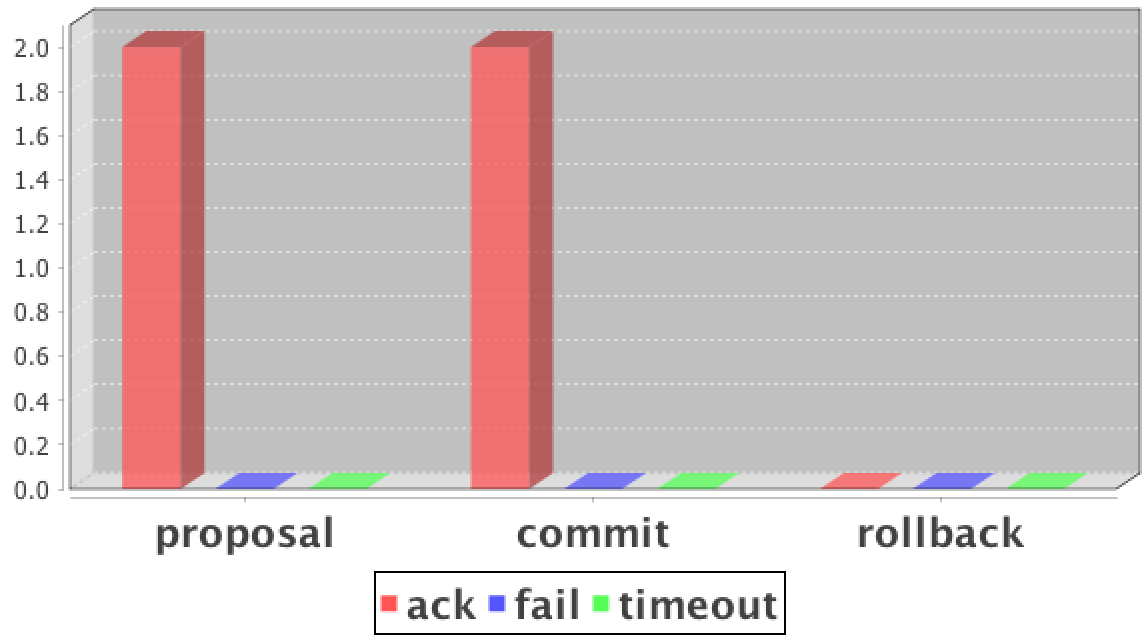}
	\label{fig:2pc_master}
}	
\subfigure[CBT Coordinator]{
	\includegraphics[width=40mm]{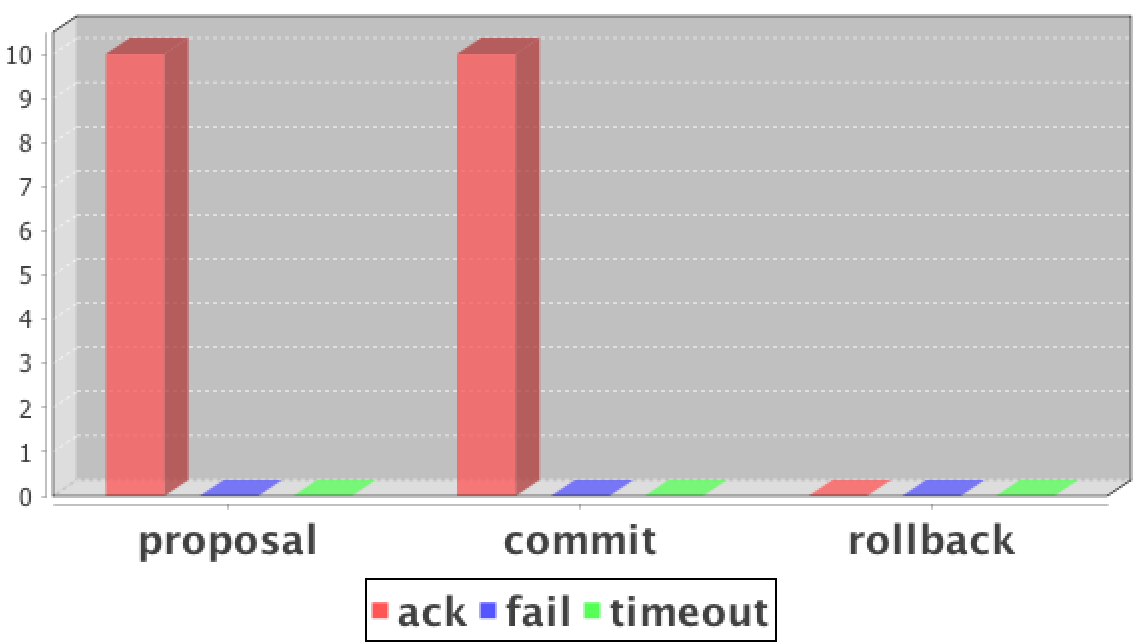}
	\label{fig:cbt_master}	
}
	\caption{Blocking 2PC vs. nonblocking CBT.}
	\label{fig:blocking_2pc_cbt}
	\figspace

\end{figure}

The first experiment verifies that CBT is nonblocking in the scenarios where 2PC is blocking due to the failures of the coordinator (between the first and second phase).
In our implementation, a successful transaction can be easily determined by checking the number of \textit{commit} messages compared with the number of participants.
Specifically, from a coordinator's perspective, every batch of $n$ commit messages implies a successful transaction where $n$ indicates the number of participants.
To verify that CBT is nonblocking as opposed to 2PC,
we send a request of five transactions to two instances of three blockchains, one with 2PC and the other with CBT.
At the end of the first transaction, we manually shut down the coordinator in both instances, and we expect the 2PC-instance will stop working on consequent transactions (i.e., blocking) and the CBT-instance will select a new coordinator from the same blockchain to continue the work.
Indeed, as illustrated in Figure~\ref{fig:blocking_2pc_cbt},
the 2PC-instance only reports one successful transaction with two commit messages (because there are two participants) to the coordinator whereas the CBT-instance coordinator reports 10 commit messages, or five successful transactions.

\begin{figure}[!t]
	\centering
	\includegraphics[width=70mm]{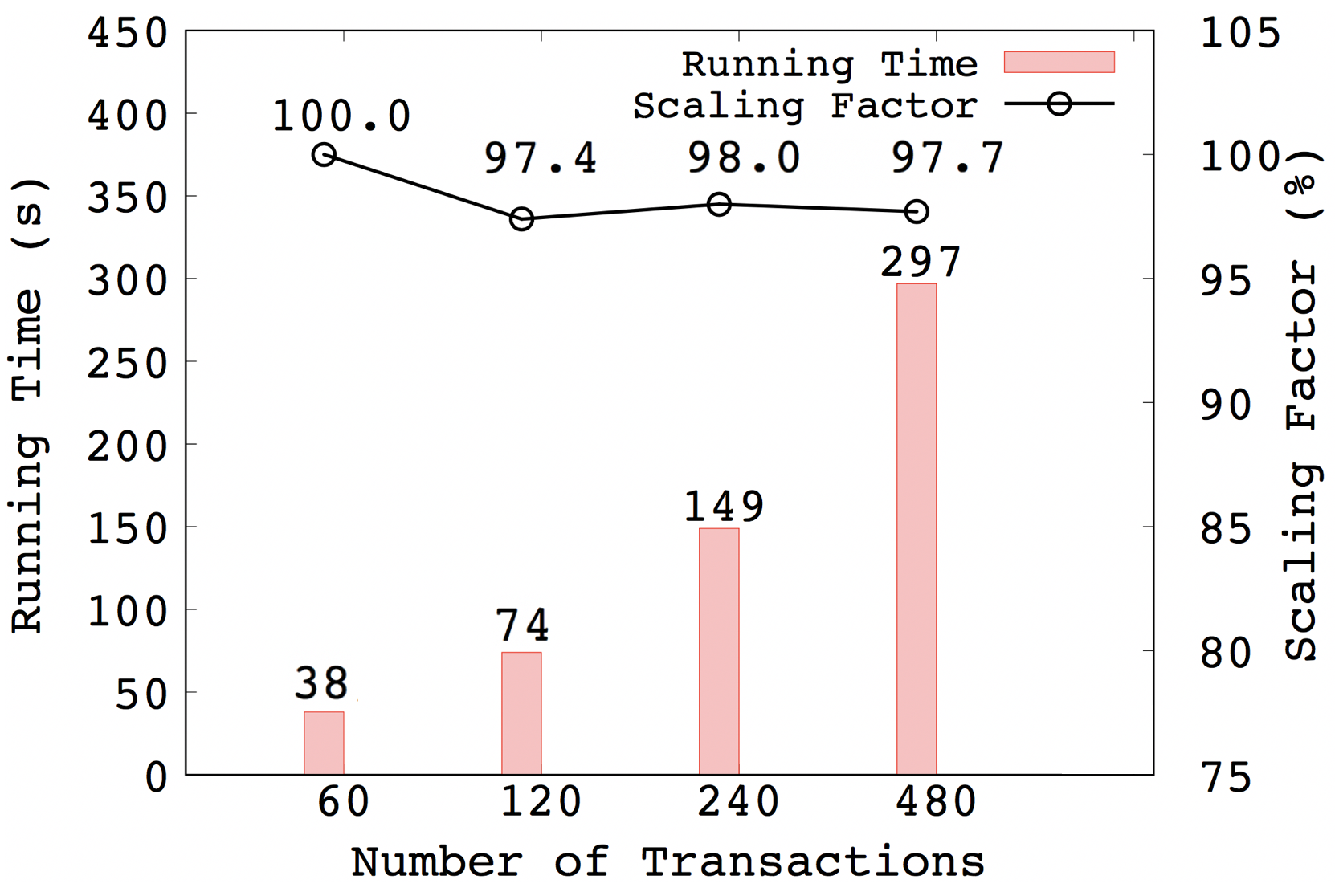}
	\caption{Different number of transactions completed by two blockchains using CBT.}
	\label{fig:diff_txns}
	\figspace
\end{figure}

The second experiment evaluates CBT's scalability:
we fixed the number of blockchains as two and submit different numbers of transactions ranging from 60 to 480.
The raw completion time is reported on the left $y$-axis in Figure~\ref{fig:diff_txns}.
The \textit{scaling factor}, defined as $\displaystyle \frac{t / t_0}{w / w_0}$, where $t$ and $w$ indicate the running time and workload at a specific scale, and $t_0$ and $w_0$ indicate both at the baseline scale.
In this experiment, for instance, $t_0 = 38$ and $w_0 = 60$.
We thus observe a closely linear scalability of CBT: 
at 120, 240, and 480 scales, the scaling factor is less than 3\% off the original, as shown on the right $y$-axis of Figure~\ref{fig:diff_txns}.

\begin{figure}[!t]
	\centering
	\includegraphics[width=70mm]{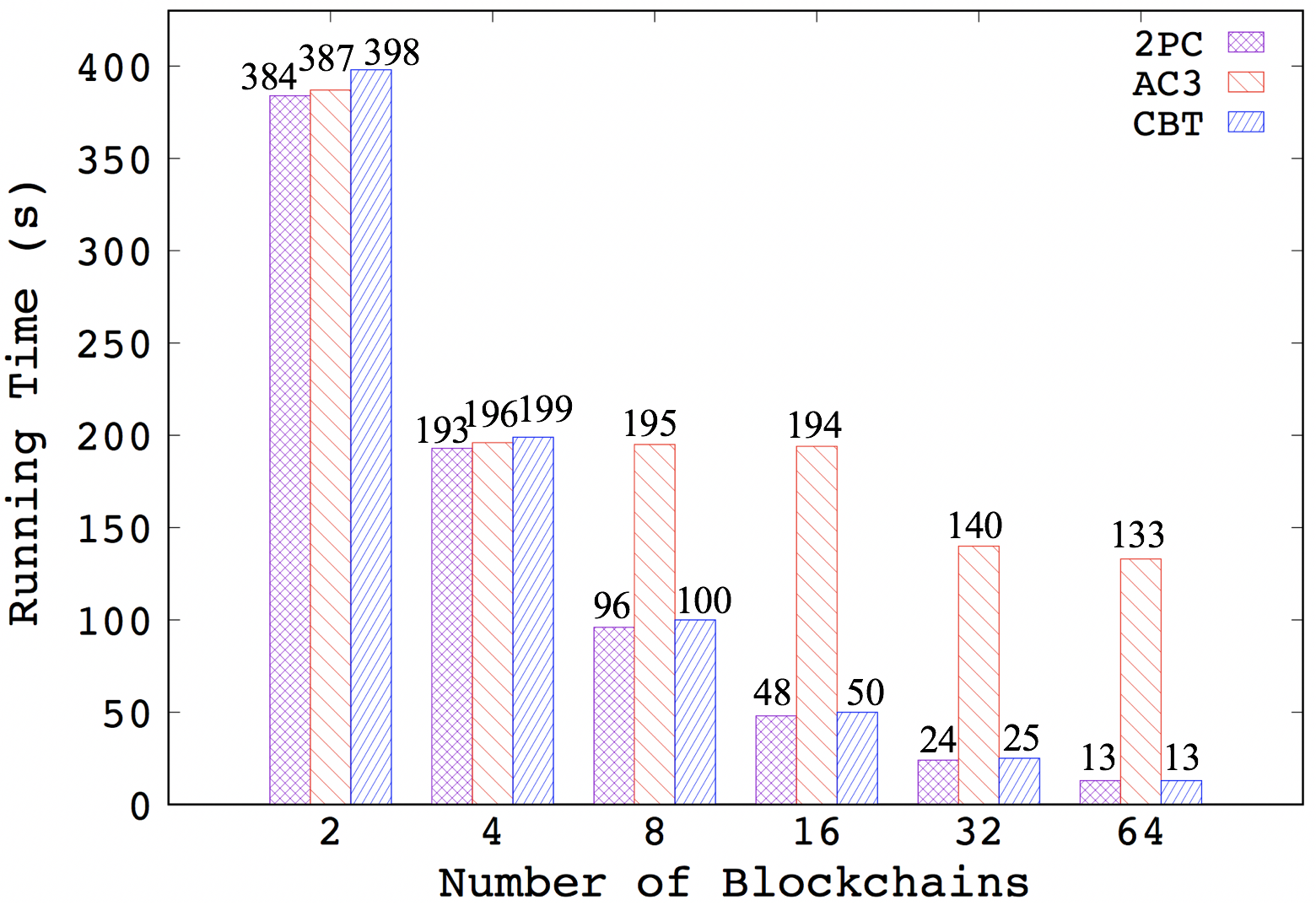}
	\caption{Running time for completing 640 transactions among different number of blockchains.}
	\label{fig:diff_workers}
	\figspace
\end{figure}

The third experiment compares CBT with two state-of-the-art protocols in cross-blockchain transactions: 2PC~\cite{2pc} and AC3~\cite{vzakhary_arxiv19}.
For 2PC, we set it up as the ``ideal case'' where no failures take place during the experiment;
it is the upper-bound performance one can best expect from 2PC.
The point is to show the overhead incurred by our proposed CBT compared to such upper-bound performance.
For AC3, we arbitrarily select one blockchain as the ``hub'', or ``witness blockchain'' as in the literature.
Because of AC3's centralized hub, we expect the performance and scalability will be affected at some point, e.g., larger number of blockchains.
We fix the workload of 640 transactions and vary the number of blockchains between 2 and 64.
The results are reported in Figure~\ref{fig:diff_workers}.
Both 2PC and CBT show (almost) linear scalability because no centralized component exists in the system.
CBT incurs insignificant overhead (compared with baseline 2PC) at small/medium scales:
3.6\% -- 4\% on 2--32 blockchains;
then the overhead is negligible on 64 blockchains.
Compared with 2PC and CBT, AC3 starts to fall behind on eight blockchains due to its ``hub'' design.
Interestingly, AC3's performance seems to follow a downward ladder trend with respect to the increasing numbers of participating blockchains:
\{2\} $\rightarrow$ \{4,8,16\} $\rightarrow$ \{32,64\}.
More investigation to this phenomena is beyond the scope of this paper and we leave this as an open question to the community.



\section{Demo Scenarios}
\label{sec:demo}

We will demonstrate our open-source implementation of various cross-blockchain transaction protocols using the a graphical user interface.
Again, the source code can be downloaded from Github:
\url{https://github.com/hpdic/cbt}.
This section will select some of the key steps in running the program.

\begin{figure}[!ht]
	\centering
	\includegraphics[width=70mm]{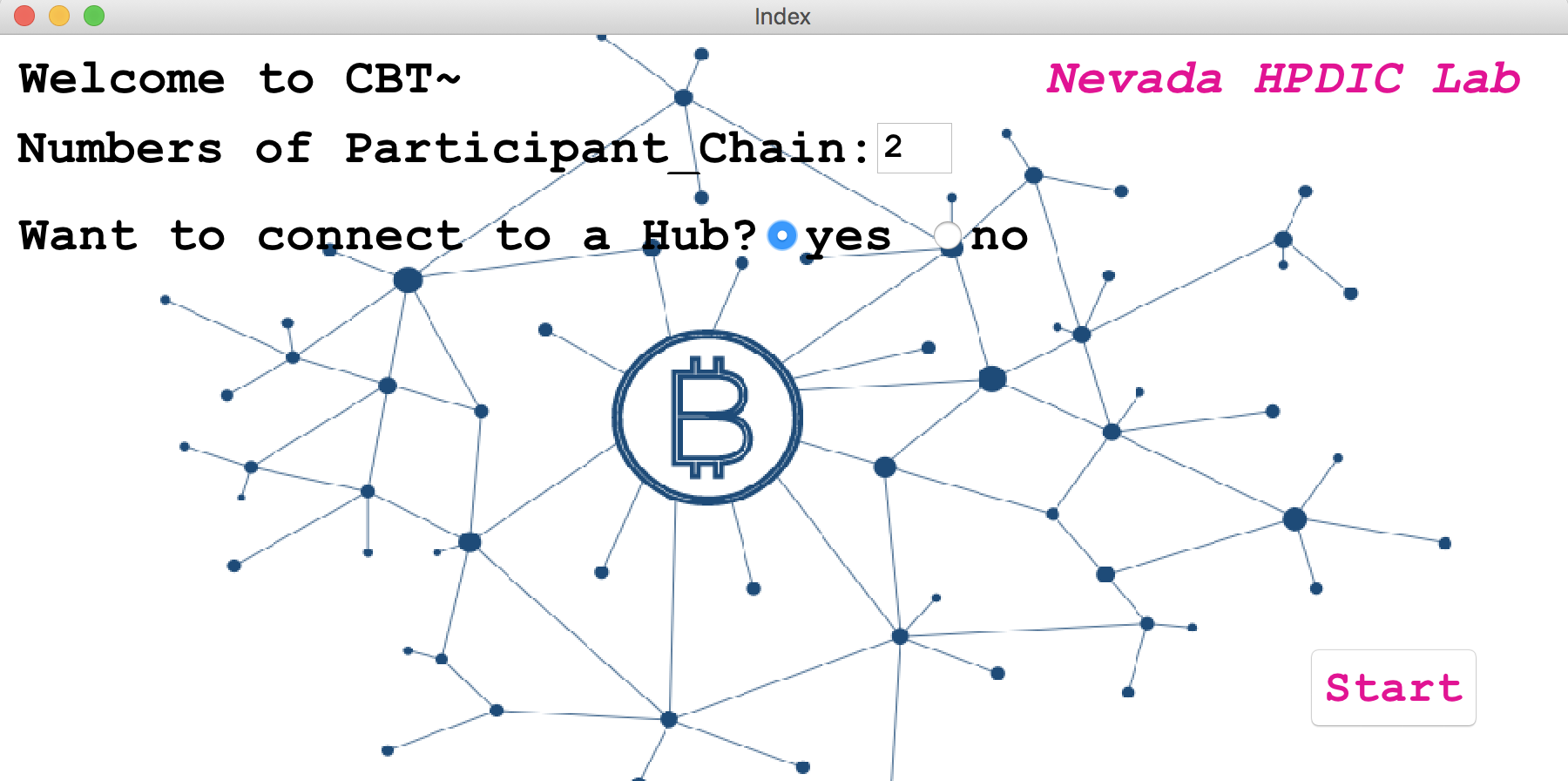}
	\caption{CBT portal.}
	\label{fig:index}
    \figspace
\end{figure}

When the user first launches CBT, she will specify the number of blockchains and whether a hub is available, as shown in Figure~\ref{fig:index}. 
For instance, here we set three blockchains: one coordinator (also the hub) and two participants.

\begin{figure}[!ht]
	\centering
	\includegraphics[width=70mm]{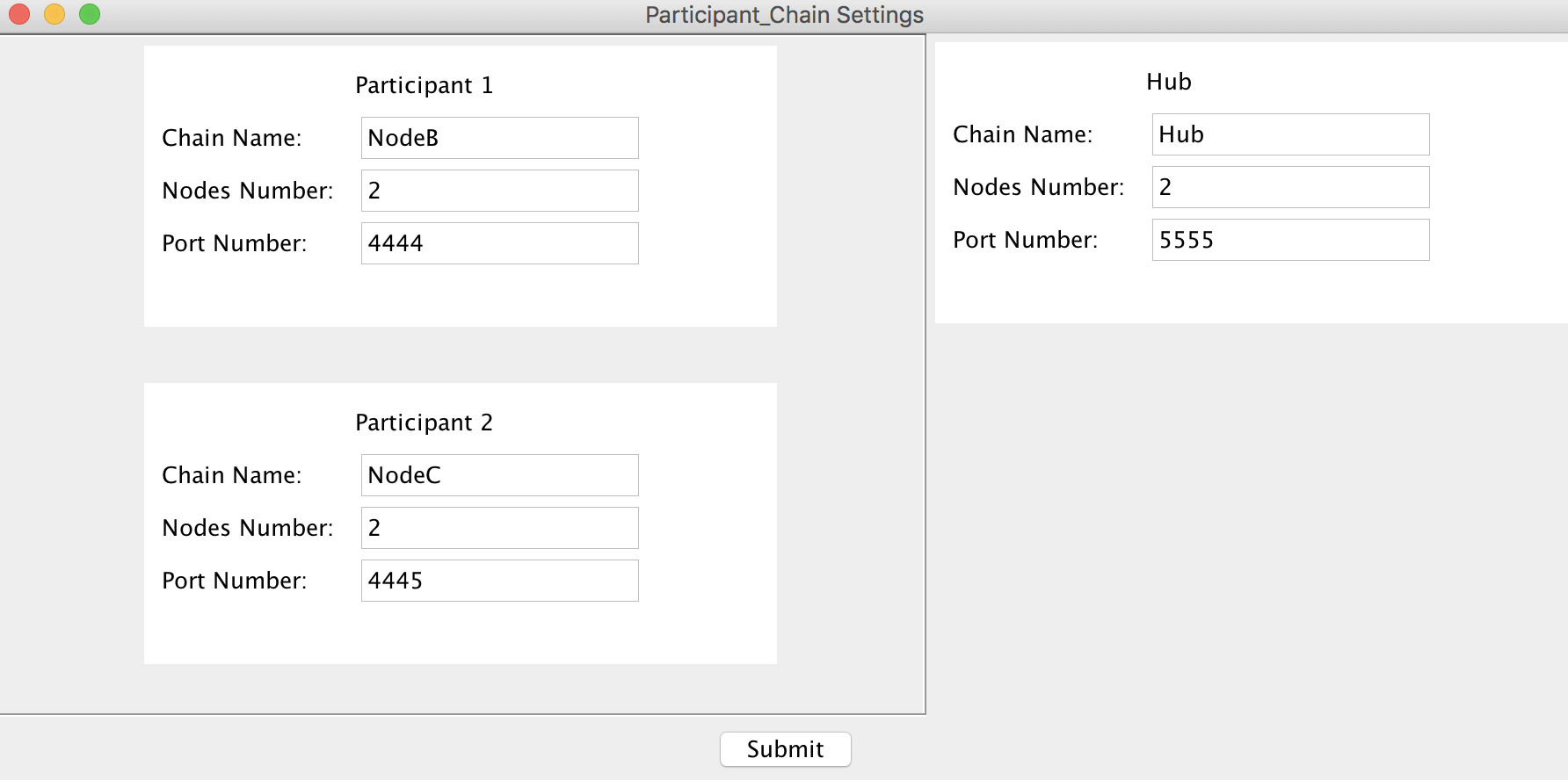}
	\caption{Parameter setting.}
	\label{fig:setting}
	\figspace
\end{figure}

Then, the user would specify the properties of each blockchain, such as number of nodes within each blockchain, the ID of the blockchain, and their port numbers.
Following our 3-blockchain instance, we will simply set two nodes for each blockchain along with other parameters,
as shown in Figure~\ref{fig:setting}.

\begin{figure}[!ht]
	\centering
	\includegraphics[width=70mm]{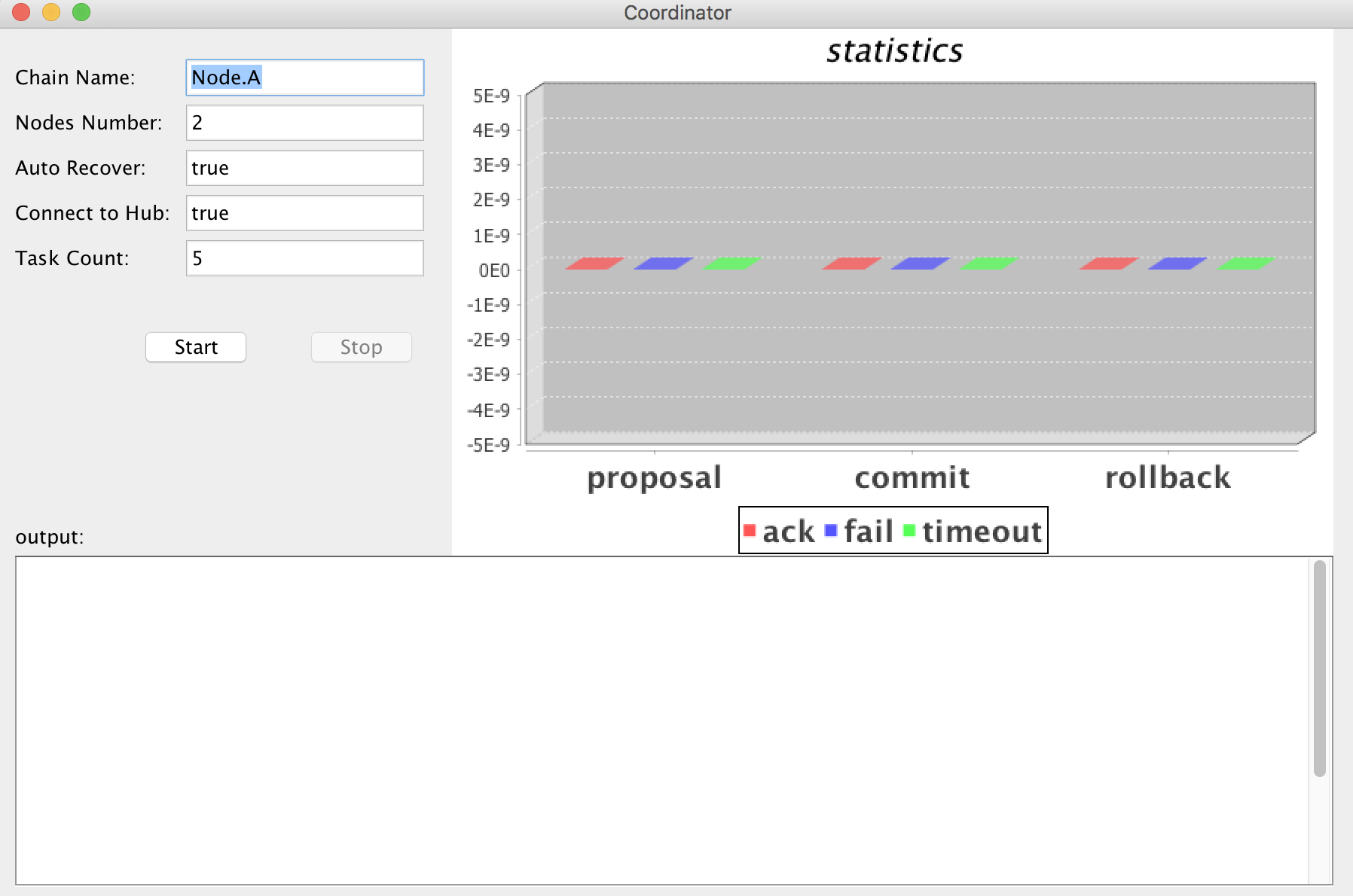}
	\caption{Coordinator panel.}
	\label{fig:coordinator}
    \figspace
\end{figure}

After the user specifies the structure of the cluster of blockchains, she can submit transactions to the system at the control panel of the coordinator, as shown in Figure~\ref{fig:coordinator}.
The screenshot shows that the user submits five transactions to a 3-blockchain cluster (on the top-left panel).
The top-right panel will report the number of messages passed during the transaction;
Figure~\ref{fig:blocking_2pc_cbt} is one such example.
The bottom panel will output detailed log information:
for instance, if one node crashes,
it will be reported here.

\section{Concluding Remark}

In the future ``full version'' of this paper, 
we will extend this work from the following perspectives.
We will provide a theoretical proof and analysis of the protocols, mainly on the safety (correctness) and liveness (nonblocking) properties.
We are currently working on scaling the CBT protocols to tens of hundreds of nodes.
Last but not least, we are working on a set of primitives, extended from SQL, to allow RDBMS users to switch to blockchain-based general-purpose data management systems backed by CBT protocols.



\end{CJK}

\bibliographystyle{abbrv}
\bibliography{ref_new}

\end{document}